# Understanding the Effects of Visualizing Missing Values on Visual Data Exploration

Hayeong Song*   Yu Fu[†]   Bahador Saket[‡]   John Stasko[§]

Georgia Institute of Technology

**ABSTRACT**

When performing data analysis, people often confront data sets containing missing values. We conducted an empirical study to understand the effect of visualizing those missing values on participants' decision-making processes while performing a visual data exploration task. More specifically, our study participants purchased a hypothetical portfolio of stocks based on a data set where some stocks had missing values for attributes such as PE ratio, beta, and EPS. The experiment used scatterplots to communicate the stock data. For one group of participants, stocks with missing values simply were not shown, while the second group saw such stocks depicted with estimated values as points with error bars. We measured participants' cognitive load involved in decision-making with data with missing values. Our results indicate that their decision-making workflow was different across two conditions.

## 1 INTRODUCTION

Data analysts often must perform analysis on imperfect datasets (e.g. failure in collection pipeline). Designers of data visualizations have faced a dilemma in how to (or how not to) represent those missing values [24]. When creating visualizations, designers have two primary options to manage missing values: 1) simply not show those data items or 2) impute new data values (calculate substitute values) based on existing data and represent the imputed values visually. Prior research in this area has largely focused on how people view and react to the representations of missing data and values in static visualizations and images [3, 10, 12, 14, 32]. That research has addressed the crucial first step in seeking to better understand how people perceive and interpret such representations. Our work builds upon this by examining how the representation (or lack thereof) of missing values affects people's interactive data analysis and decision-making process.

*Our research seeks to first, understand how people make decisions when data contains missing values. Second, how visualizing missing values in a dataset affects the visual data exploration process.* We hypothesize that imputing and representing missing values could help people to better consider the potential impact of absent data values by enabling them to reason about the quality of their dataset during the visual data exploration process. Also, we hypothesize that this would increase the cognitive workload to users.

We conducted a formative empirical study to understand steps people take when they have to make decisions with a dataset that contain missing values based on our previous work [31]. Participants were presented with a stock buying scenario. To explore and analyze stock data, participants used an interactive scatterplot tool. The study contained two conditions. One group of participants performed data exploration aided by a scatterplot visualization that did not present (hid) data items with missing values. The second group of participants performed the same task using a scatterplot visualization that imputed missing values and represented them in the visualization.

Our findings examined the steps that people took to make decisions when presented with a dataset that contains missing values. Also, our findings show that whether or not one visually represents missing values impacts participants' decision-making workflow. We observed that when missing values were visually represented, participants used those visual representations to reason about data quality, and their decision-making processes exhibited greater consistency and regularity. The participants also reported more confidence in their responses.

## 2 RELATED WORK

In our work, we define missing values in data as those attribute values that are missing for data items in a multivariate data set. We do not focus on the reason or cause of the missing values, just their (lack of) existence.

### 2.1 Analyzing Incomplete Data

Missing data values can occur throughout the data life cycle and if missing values are not cleansed properly it can result in unreliable analysis [27]. Therefore, when missing values exist, these values are often cleansed, extracted, and integrated as part of the data wrangling process to suit the context of data being used for data exploration and analysis [18]. Along those lines, many visualization systems support data quality analysis such as by showing data transformation history over time [7] or some systems such as Wrangler [25] support automatic inference to handle missing values for analysts to transform data. Wong & Varga [36] suggest to indicate missing values in order to help analysts know where data are missing help them to reason about data quality. Visualization systems indicate the presence of missing values to help with data quality analysis [5, 6, 17]. For example, xGobi [33], MANET [35], and VIM [34] indicate the amount of missing values. Based on this prior work, for one of the conditions that we do not visually represent missing values, we elect to indicate that missing values exist in the dataset and indicate the number of missing values.

To indicate where data values are absent, missing values can be approximated and visually represented with different types of marks. These different types of marks are often drawn from uncertainty visualization methods because imputed values present uncertain information. In prior studies of uncertainty visualization, a variety of visualization features (e.g., salience [9], sketchiness [37]) were manipulated to represent missing values. Other visual techniques such as error bars [8, 10, 20, 26, 28, 32] have been used to indicate missing values.

### 2.2 Evaluating the Impact of Visualizing Missing Values

Prior work tells us that whether or not missing values were visually represented changes the way people interpret data. Methods of visual representation can impact people's perception of data quality and confidence level in their response [2, 8, 10, 21, 26, 28]. In addition to changing people's perception of data quality and confidence, methods for visualizing missing values can impact the decision-making process (e.g., effectiveness of uncertainty visualization in game-like tasks [4] or transit decision making tasks [16, 26]).

*e-mail:hsong300@gatech.edu
[†]e-mail:fuyu@gatech.edu
[‡]e-mail:saket@gatech.edu
[§]e-mail:john.stasko@cc.gatech.edu

A body of prior work measures the impact of visualization choices on statistical analysis. Eaton et al. [13] measured the effects of visualization choices on trend estimation tasks. Study results showed that participants preferred missing values that were indicated explicitly. Song & Szafir [32] measured the effects of imputation methods and visualization methods (e.g., highlighting) on perceived data quality in time-series data. Results showed that information removal, not representing missing values, degraded the participants' perception of data quality and confidence and even led incorrect responses of participants on statistical comparison and trend detection task.

Prior work evaluated the impact of visualizing missing values on low-level tasks (e.g. trend detection) and static visualization, which led to conceptual findings. Our work extends these conceptual findings in graphical perception to evaluate the impact of visualizing missing values in people's visual data exploration by employing interactivity and a data-driven task. We employ a scatterplot as a representation because it helps us determine how people understand patterns (e.g. a correlation between attributes).

## 3 User Interface: Interactive Scatterplot

To investigate how people explore and make decisions when a data set is incomplete, we built an interactive scatterplot visualization.

Within the scatterplot visualization, each item in the dataset is represented by a small circle. The system supports two versions of the scatterplot that handles the presence of missing values.

**Baseline.** The first version does not visually represent a data item in the view if it is missing a value for either of the two attributes actively being displayed. Additionally, text in the upper right indicates how many data items are missing from the view currently.

**Error bars.** The second version imputes visually represents data items that are missing one of the displayed attributes (Figure 1 depicts the system interface). To impute a missing value, we used k-nearest neighbors (three) to estimate the missing value, in which all the other remaining attributes in the dataset as features to find most similar data items and computed average values based on those data items. A bar's length is the standard deviation value. A horizontal line indicates that the item's attribute on the x-axis is estimated, and a vertical line indicates that the attribute on the y-axis is estimated.

Many existing visualization systems such as Tableau do not necessarily visualize missing values but rather indicate the number of missing values and give option for users to remove those values. Inspired from this approach, we wanted a baseline condition that explicitly did not represent a data item with a missing value in any way but indicate the number of missing values through text.

## 4 Study Design

We conducted a between-subjects experiment to study the influence of representing or not representing missing values on people's decision-making and visual data analysis process. The study contained two conditions: *baseline* and *error bars*. We randomly divided the 18 participants into two groups, 9 participants in each group (P1-P7, P9, P16: *error-bars*, P8, P10-P15, P17-P18: *baseline*). We measured how representing missing data affects the level of mental demand and frustration to participants during the data exploration process. The study took about 1 hour to complete and we compensated participants with a $10 Amazon gift card. Each session followed the same general procedure. Study materials are available in the supplemental materials.

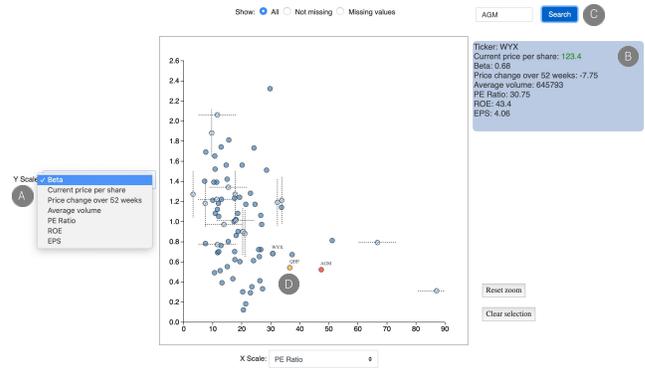

Figure 1: Here, we show the version of the system in which missing data are visually represented. (a) Users can assign data attributes to the x and y axes using the drop down menu next to the screen view. (b) Users can hover the mouse cursor over a data item (stock WYX here) to see all of its data attribute values. (c) Users can search for an item by name (highlighted in red). (d) Users can track a data item by clicking on its circle, which colors it orange and adds a label to it. These interactions were selected based on the interaction framework by users' intent [38] to support visual exploration for people.

**Benchmark Dataset.** For a stock buying scenario, we first scraped stock information from Yahoo Finance [1] for 81 different companies and 8 attributes for each stock: ticker symbol, ROE (Return on Equity), EPS (Earnings Per Share), price change over 52 weeks, current price per share, average volume, beta, and PE ratio (Price-to-Earnings ratio). The ticker symbol is a text value and all the remaining 7 attributes are numerical values. We anonymized company names and ticker symbols in the dataset to prevent participants from making decisions based on their prior knowledge about a specific company. The initial dataset that we scraped was complete and did not contain missing values. To make the dataset incomplete, for each attribute we randomly removed about an equal number of low, medium, and high valued attributes. Note that in practice missing values are less often a random selection of attributes but instead follow some regular pattern. For example, missing data might have a regular pattern due to data transmission error [27]. Schouten's work [30] explores methods for removing values along regular patterns.

**User Task.** To observe participants' decision-making processes, we created a task that requires them to reflect on the data and consider the variety of attributes per item in the data set. Their task would be, given an initial pot of money to begin, to construct the best possible stock portfolio for the future. We asked participants to verbalize their thoughts while exploring and making decisions, and we tried to avoid interrupting the participants as much as possible during their data exploration and analysis process. We encouraged participants to spend as much time as needed, and we allowed them to use the tutorial throughout the study session.

**Survey & Interview.** When each participant finished the task, we interviewed them about their experiences. We created a list of questions to understand their decision-making process, the influence of the visualization, and their user experience. We asked participants to self-report mental demand and frustration level (NASA TLX-survey [23]), to understand the effects of visual representation of missing values on cognitive load in decision-making. We also sought ideas for improving the system.

## 5 Results

**Survey: Mental Demand & Frustration.** To measure cognitive load involved in the data analysis process, we used frustration level and mental demand as metrics [23]. We asked participants to rate mental demand and frustration levels while performing the experimental task ( 10-point Likert scale, 1-very low and 10-very high). Participants in general reported high levels of mental demand to perform the task, but there was no significant difference (t (14.53) = -1.11, p = 0.85) across conditions (*baseline* (M = 8.22, SD = 1.39) and *error-bars* (M = 7.33, SD = 1.96)). There also was no significant difference in participants' frustration level (t (15.39) = -0.1, p = 0.54) across conditions (*baseline* (M = 5, SD = 2.06) and *error-bars* (M = 4.88, SD = 2.52))

## 5.1 Data Analysis: Data Exploration

**Video & Interview Data.** We recorded the screen and audio during the user study and the follow-up interview. To characterize and understand how participants make decisions when missing values are visually represented or not, we employed a think-aloud protocol to understand their mental process [15]. To analyze the video and interview material, we employed Creswell's qualitative data analysis approach [11].

We performed two coders evaluations (first and second author). Two coders coded qualitative data with a special focus on participants' *decision-making workflow* (What were the initial steps participants took to make decisions? What approach did participants' take to make judgments about the data item that contained missing values? We both analyzed a full collection of corpus and generated a codebook. We identified frequently occurring codes to form higher-level descriptions and narratives. Then we iterated this process until we reached a reasonable agreement, which resulted in the removal of three codes and two additional codes. This resulted in Cohen's Kappa score for visualization 0.91 (perfect agreement). Once we agreed on a code-book the first-author re-coded qualitative data.

Our video and interview analysis revealed that the participants' decision-making workflow was different across the two conditions. We characterized different actions participants took and summarized those into six steps that participants typically followed. The two sequences are summarized in Figure 2. The high-level structure of the decision-making workflow has much in common with the well-known Pirolli-Card sense-making loop [29].

### 5.1.1 General Decision-Making Workflow

**1. Strategy.** Participants first developed strategies in order to make their decisions. They determined which attribute values to focus on and established a set of criteria to help look for appropriate stocks (e.g. positive price change over 52 weeks and beta over 1.0).
**2. Explore.** Participants explored the dataset through the view to gather information and searched for stocks that meet those criteria.
**3. Identify Alternatives.** Participants identified alternative stocks that mostly met their criteria (some with or without missing values)
**4. Weigh the Evidence.** After they identified alternatives, participants weighed the evidence to decide if the selected alternatives are worth investing in for the future. They compared these stocks with other data items to cross-check and judge if the stocks likely would return with future profits.
**5. Choose Among Alternatives.** After participants made those judgments, they selected stocks from the identified alternatives for their final portfolio.
**6. Review the Selection.** Lastly, participants reviewed their selections.

Based on our observations, we found key differences across the two conditions in the phases **weigh the evidence**, **strategy**, and **review the selection**.

**Weigh the evidence.** In **error bars** condition, *participants used visual representations to weigh the evidence of data items for future selection*. For example, if the selected alternatives contained missing values, they used the visual representation to reason about the data quality of estimation in relation to other data items for future selection. Participants looked at the overall trend in the view, and if the estimation followed the trend, they judged that it was reasonable and decided to keep it. In this case, they moved forward to the next phase (see Figure 2-A). In **baseline** condition, *participants struggled to weigh the evidence and because of this they often became entangled in that decision-making process*. Participants reported difficulty to **weigh the data evidence** if the data item contained missing values, for two reasons. First, participants reported difficulty figuring out which stocks were missing and which attributes were missing in stocks to weigh the evidence (6/9 participants reported). The second reported challenge was that the participants had difficulty to compare and cross-check with other data items for future selection because of the lack of information (see Figure 2-B). All of the participants in this condition expressed in some way this difficulty of comparing data items to decide for future selection. For example, *P15* commented, *"If it [a stock] had one variable missing, the one that I was looking for, it was hard to cross-check against different variables. ... If I didn't have something to cross-check against it, I was not sure how to act when one of these crucial points was missing."*
**Strategy.** In **error bars** condition, *we observed that participants tended to follow their original strategy. Consistency and regularity stood out in participants' decision-making processes.* For example, if they decided to drop a data item with a missing value, they used the visualization to find similar alternatives, which still aligned with their initial strategy. *P13* said, *"[After knowing that the data item he was interested was missing critical information and decided to drop it], I instead looked for one that seemed similar that did not have that missing [information]......[I looked at the visualization] I considered every metric and looked at those [a cluster] to look for the similar ones"*. In **baseline** condition, *participants often changed their strategy, because they could not perform intended comparisons with data items with missing values. This deviated from their original plan*. When participants struggled to **weigh the evidence** for data items with missing values, they often became entangled in that decision-making process instead of moving forward, which made it challenging to follow their initial strategy. For example, *P8* stated, *"If I noticed that there were missing values, missing stocks, on the graph, I could not make some intended comparison using that graph. ... Then I either had to ignore [drop it] or start over again [go back and look for other attribute values to focus on]. It was difficult to stick with a plan [strategy]."*
**Review the selection**. Because of the prior differences in their analysis processes, participants reported that they felt different about their decisions. In **error bars** condition, when participants reviewed their selections, they reported confidence in their decisions. For example, *P5* explained, *"I was able to control for the missing values in the sense that whether I decided to include a missing value or not, I know that was a conscious decision on my part. I didn't accidentally include it simply because I didn't know about it. ... I made a choice to minimize the unknowns. This is reassuring to me and it gives me greater confidence in my choices.* In **baseline** condition, we did not find comparable evidence that participants felt confident about their decisions. *P18* expressed a common view that, *"I was not sure whether I did the right thing [made the right selections]."*

## 5.2 User Feedback

**Visualize to what extent?** We found differences in the level of detail participants desired to indicate the presence of missing values in a dataset. In the interviews, we asked participants to suggest potential improvements to the current way the application handles missing values across the two conditions.

Within **both** conditions, five participants (two *error-bars* and three *baseline*) expressed a desire to see details of the missing information on request rather than as a default. P4 said, *"At the very front I don't want to see it, because just knowing that those have missing values are enough for me. I would prefer to see the estimations only when I'm interested in the data item, otherwise just knowing that it is a missing value is fine for me."*

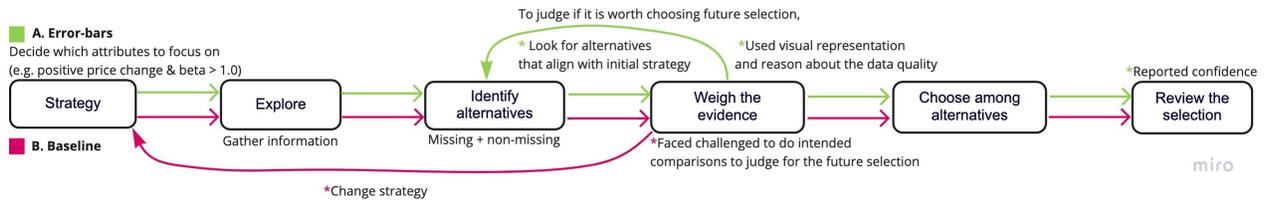

Figure 2: Stages of the observed decision making process in *error-bars* (A) and *baseline* (B). The results revealed that in condition (A), participants were more likely to stick with their original plan. Their decision-making process was consistent and regular because they were able to weigh the evidence for data items with missing values. In (B), participants faced challenges when attempting to weigh the evidence to select data items for future selection. These challenges were that it was harder for participants to locate where a data item is missing and it was difficult to compare such a data item with other data items. This tended to lead participants to be entangled in that phase and they had to regenerate a new strategy.

In the *error-bars* condition, some participants wanted specific information about the estimation quality or methodology. For instance, four participants stated that it would have been helpful if the range that they saw had been denoted by a 95% confidence interval. P6 said that *"I think adding the definition [estimation quality], which is the 95% confidence interval, will be helpful. Then I'll be 95% is a pretty good estimate. Then I am fairly certain that it falls within this range, and then maybe I would be more willing to pick something that falls in that range."* In the current application, we do not specify the quality (e.g. 95% confidence interval) of the estimation presented with error-bars.

## 6 DISCUSSION

Our work identified different steps people took to make decisions with a dataset that contains missing values. Also, our findings indicate key differences between the two conditions of our study in the participants' decision-making workflow. From the empirical observations we found the following:

- The representation of missing data values led to a change in the participants' decision-making workflow.
- We did not find significant difference in cognitive workload across two conditions.
- In the *error-bars* condition,
  - Participants were more likely to stick with their original strategy.
  - Consistency and regularity stood out in their decision-making process.
  - Participants reported confidence in their decisions.

**Support for Decision-Making Process.** Our results indicated that *visually representing missing values can assist participants to reason about data* and can change their decision-making process. We observed that visually representing missing values helped the participants to focus attention to the missing values when making decisions. This somewhat aligns with the impact of visual selection [19], in which missing values were visually represented with different visual marks and error bars. This focus in attention to missing values encouraged the participants to make judgments about the data quality and to decide whether the missing values that were visually represented were trustworthy or not.

Based on these judgments, participants chose alternatives and followed their strategies while incorporating the uncertainty involved in the dataset. Participants understood the overall trends of the data by identifying correlations between two attributes and observing characteristics in the data (e.g., clusters and outliers). For example, if the data item broke the continuity of the trend (graph), then they perceived the quality of that data item to be lower. This aligns with findings from Song & Szafir [32], when the visual representations of missing values preserved the continuity of a graph, people perceived the data quality to be higher, and vice versa. These comparisons and analyses helped the participants to make judgments about the data quality and develop their strategy, whether or not to add a newly selected stock to their portfolio. Because the participants were able to make judgments about data quality, they reported confidence in their responses.

**Opportunities for Future Research.** If we are designing a visualization system that can assist people to make decisions, with an incomplete dataset one could consider designing when users are in these different phases. For example, some participants did not want to see missing values up-front when they were *exploring dataset at the beginning* but wanted to see information about missing values up-on demand when they *scoped down their selection*. Also, when people are scoping down their selection, participants wanted information about severity of the missing values and imputation quality so that they could weigh the evidence to make their final selection. Based on these findings, we see value in designing a visualization system that can communicate uncertainty of dataset to people in different decision-making phases as a future research direction, such as supporting event sequence predictions [22] in a system.

## 7 LIMITATIONS AND FUTURE WORK

A limitation of the study involves the ecological validity of the stock buying scenario. Typically, investors making stock purchases have additional information available such as the actual stock name, historical performance data, market analysis, etc. So in many ways, our analysis task was a simplified version of the process that someone might typically conduct with this data. Also, our participants were graduate students who do not represent a broad cross-section of the populace. Further, these students typically do not buy stock portfolios and thus they are less familiar with the domain of the study than would be investment professionals. Finally, the study included a relatively small number of participants so one should be careful not to overly generalizes the findings.

## 8 CONCLUSION

In this work, we observed how people explore and make decisions when missing values were visually presented or not. We performed a user study in which participants selected portfolios of stocks, some with missing attribute values. We found that the presence or absence of missing value representations changed the participants' decision-making process. Because of these differences, participants' perceptions of their decisions were different. The participants felt more confident about their decisions when missing values were visually represented. We found preliminary results for how visually representing missing values can help people's data exploration and decision-making.